\documentclass{article}

\PassOptionsToPackage{numbers, compress}{natbib}

\usepackage[preprint]{neurips_2026}

\usepackage[utf8]{inputenc} 
\usepackage[T1]{fontenc}    
\usepackage{hyperref}       
\usepackage{url}            
\usepackage{booktabs}       
\usepackage{amsfonts}       
\usepackage{nicefrac}       
\usepackage{microtype}      
\usepackage{xcolor}         
\usepackage{graphicx} 
\usepackage{subcaption} 
\usepackage{amsmath}
\usepackage[table]{xcolor}
\usepackage{makecell}
\usepackage{multirow}
\usepackage{float}

\usepackage{algorithm}
\usepackage{algpseudocode}

\usepackage[most]{tcolorbox}
\usepackage{enumitem}
\usepackage{booktabs}

\usepackage{multirow}

\usepackage{tabularx}

\newtcolorbox{promptbox}[1]{
  title=\textbf{#1.},
  colback=gray!4,
  colframe=gray!99,
  boxrule=2.5pt,
  arc=2pt,
  left=6pt,
  right=6pt,
  top=6pt,
  bottom=6pt,
  breakable,
  enhanced,
}
\title{Protecting Shared Agent Workflows from Invalidation Attacks with Signed Trajectories and Runtime Verification}

\author{%
  {Ruixiao Li} \\
  School of Cyber Science and Engineering\\
  Xi'an Jiaotong University\\
  \texttt{lrx4536257@stu.xjtu.edu.cn} \\
  \And
  Fahao Chen \\
  School of Artificial Intelligence\\
  Shandong University\\
  \texttt{chenfh@ieee.org} \\
  \And
  Peng Li \\
  School of Cyber Science and Engineering\\
  Xi'an Jiaotong University\\
  \texttt{pengli@xjtu.edu.cn} \\
}

\begin{document}

\maketitle

\begin{abstract}
Large language model (LLM)-based agents increasingly rely on shared workflow repositories to transfer experience across tasks and agents. By abstracting past executions into reusable workflow graphs, such systems significantly improve planning efficiency and generalization. However, this emerging paradigm introduces a new and underexplored security risk: if the shared repository is compromised, tampered workflows can systematically mislead downstream agents, inducing persistent execution failures or unsafe behaviors.
In this work, we present a principled protection framework for shared agent workflows based on signed trajectories and runtime verification. We require all cloud-collected trajectories used for workflow generation to be cryptographically signed, with each trajectory representing a concrete execution instance. Prior to repository inclusion, trajectories undergo a unified screening process to ensure correctness and validity. When an agent retrieves a workflow from the shared repository, it also obtains a small set of signed supporting trajectories and verifies both their cryptographic integrity and their semantic consistency with the workflow. To make this practical for branching workflows, we further propose a runtime, path-aware retrieval strategy that dynamically fetches only the signed trajectories relevant to the execution path actually taken by the agent, substantially reducing communication and verification overhead. 
Extensive experiments across multiple reasoning benchmarks show that workflow invalidation attacks achieve up to 90.12\% success rates, degrading agent inference performance by up to 73.16 percentage points in accuracy without effective workflow protection.
In addition, the proposed protection achieves an average F1 score of 85.26\% and a detection rate of 79.97\%, outperforming existing baselines by about 2.02×.

\end{abstract}

\section{Introduction}
\label{submission}
\begin{figure*}[t]
\vspace{-4pt}
\centering
\centerline{\includegraphics[width=\textwidth]{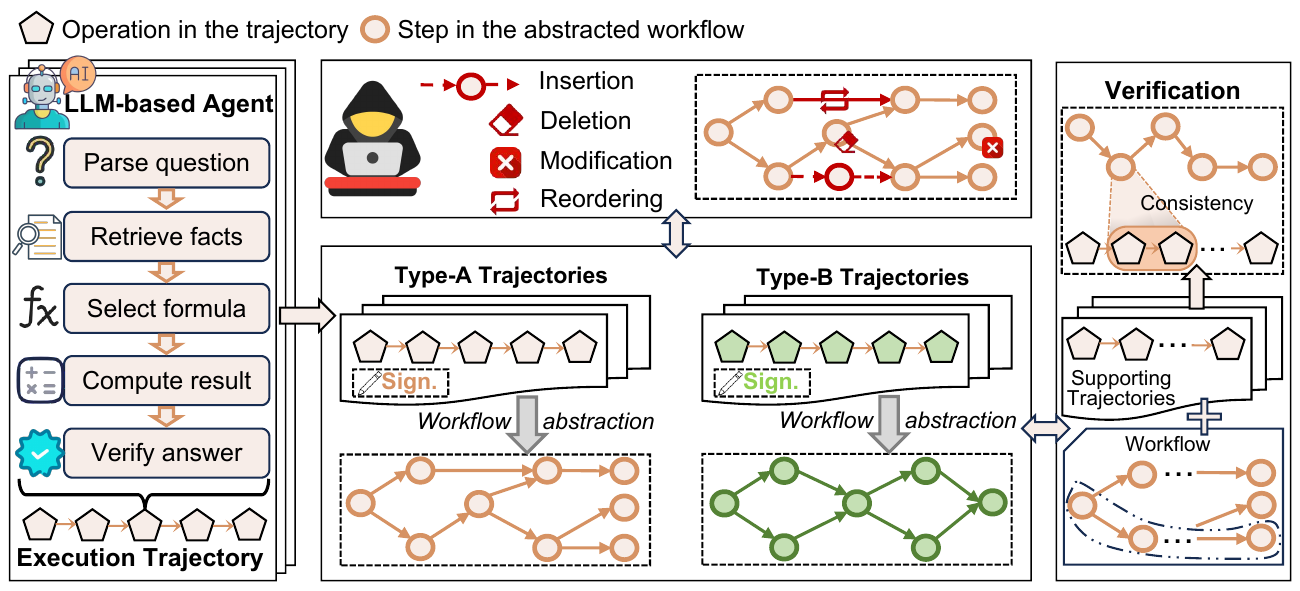}}
\caption{ (\textit{\textbf{Left}}) LLM-based agents upload signed execution trajectories to a cloud repository, where trajectories are aggregated and abstracted into reusable workflows (\textit{\textbf{Center-Bottom}}). A compromised repository can stealthily invalidate workflows by modifying their structure or execution paths while preserving plausibility (\textit{\textbf{Top}}). Our framework mitigates such attacks by grounding workflows in authenticated trajectories and enforcing path-aware verification (\textit{\textbf{Right}}).}
\label{introduction}
\vspace{-16pt}
\end{figure*}

Large Language Model (LLM)-based agents have recently become a powerful foundation for building autonomous systems capable of perceiving, reasoning, planning, and acting in complex environments \cite{yao2022react}. Modern agents increasingly rely on accumulated experience, such as historical trajectories, to improve robustness and efficiency across tasks \cite{zhong2024memorybank,schick2023toolformer,wu2025evolver}. To make such experience transferable, recent work has proposed abstracting raw execution trajectories into high-level workflow-structured representations that capture the essential states, actions, and control logic underlying successful task executions \cite{wangagent,zhao20252flow}. These workflows serve as reusable experience units that can guide downstream agents when solving new but related tasks, significantly improving planning stability, reducing exploration cost, and enhancing generalization \cite{liu2024skillact}.

In realistic deployments, an individual agent rarely accumulates sufficient trajectories to extract reliable workflows. This has led to the widespread adoption of shared workflow frameworks, where a cloud aggregates trajectories from many agents, synthesizes workflows using large models, and maintains a shared workflow repository \cite{yang2024cops,feng2025get,zhang2025g,tang2025agent,zhang2025survey}. When encountering unfamiliar tasks, an agent queries this repository to retrieve relevant workflows and incorporates them into its planning process \cite{gao2024memory}. 

While effective, this architecture introduces a critical yet underexplored security vulnerability: the shared repository becomes a high-value target. By design, it aggregates trajectories from many agents and directly supplies decision-critical experience to all downstream users. As shown in Figure~\ref{introduction} \textit{Top}, an adversary may compromise agents prior to trajectory submission or infiltrate the centralized service after repository construction, enabling large-scale, direct or indirect tampering with trajectories and workflows. Such manipulation can stealthily influence the behavior of a broad population of agents. This attack surface is known to be highly impactful in centralized learning and data aggregation systems
 \cite{biggio2012poisoning,steinhardt2017certified,bagdasaryan2020backdoor,koh2021stronger}. In this work, we term this class of attacks \emph{workflow invalidation attacks}.

We assume that most trajectory-submitting agents follow the protocol (Figure~\ref{introduction} \textit{Left}), while a small fraction may be compromised and submit tampered trajectories, consistent with common settings in shared or collaborative learning frameworks \cite{bonawitz2017practical,kairouz2021advances}. If the repository aggregates such corrupted inputs or is further compromised, downstream agents may be systematically misled during planning and execution. Because workflows directly shape long-horizon action sequences, even subtle modifications can induce persistent execution errors, violate safety constraints, and potentially cause severe real-world consequences. Workflow invalidation attacks are fundamentally different from prompt-level or observation-level attacks: rather than targeting individual interactions, they exploit the shared experience layer that many agents trust by design, enabling scalable and persistent influence over agent behavior.

To defend against workflow invalidation attacks, we propose a lightweight yet principled protection mechanism based on trajectory signing and verification. Concretely, agents that submit trajectories to the shared repository are required to cryptographically sign their trajectories. Upon collection, the server performs unified screening via cross-validation and embedding-based similarity to identify and filter out anomalous or potentially corrupted trajectories. The remaining signed trajectories are used both to construct workflows and to serve as verifiable evidence. The server stores both workflows and their associated screened, signed trajectories (Figure~\ref{introduction} \textit{Center-Bottom}). As shown in Figure~\ref{introduction} \textit{Right}, when an agent downloads a workflow, it also requests several signed trajectories that were used to derive or support this workflow. By verifying the signatures, the agent can ensure that these trajectories have not been modified after submission. The agent checks whether the retrieved workflow steps remain semantically consistent with the authenticated trajectories. Any unauthorized modification, either to the trajectories or to the derived workflows, can therefore be detected. This design shifts trust from the centralized repository to a distributed provenance mechanism, enabling agents to validate the integrity of shared experience even if the cloud is partially compromised.

Implementing this this mechanism raises a key challenge: Workflows are typically structured as directed acyclic graphs with multiple possible execution paths, while a single trajectory only instantiates one concrete path. Naively downloading all signed trajectories corresponding to all possible paths would incur prohibitive communication and storage overhead. We therefore propose a runtime, path-aware retrieval strategy. Instead of prefetching all evidence, the agent dynamically requests only the signed trajectories corresponding to the path it actually executes. When the agent’s execution reaches a branching point, it selects the next branch based on the current environment state and then downloads the associated signed trajectories for that branch. This design ensures authenticated evidence for the executed path while avoiding irrelevant trajectory transfer.

\textbf{Evaluation.} We conduct extensive experiments on multiple widely adopted and synthetic datasets. Results show that workflow invalidation attacks can reduce agent accuracy by up to 73.16 percentage points. Our method effectively mitigates these attacks,  achieving a detection rate of 79.97\% and outperforming existing baselines by an average factor of 2.02$\times$.

\section{Background and Related Works}

\textbf{LLM-based agent.} LLM-based agents augment LLMs with auxiliary functional modules, enabling them to solve complex tasks beyond pure language reasoning~\cite{wang2024survey,cai2025building}. In such systems, the core LLM typically handles high-level reasoning, planning, and decision-making~\cite{yao2022react}, while auxiliary modules, such as external tools~\cite{schick2023toolformer, patil2024gorilla} and memory or experience components~\cite{xu2025mem}, allow agents to interact with real or simulated environments~\cite{shridhar2021alfworld} and accumulate long-term experience~\cite{zhang2025g}.

\textbf{Agent workflow.} As experience accumulation and reuse become increasingly central to agent systems \cite{liu2025advances,park2023generative}, effective representation and organization of experience are critical to agent performance and generalization~\cite{wang2022selfconsistency}. Beyond unstructured summaries, recent work has proposed representing experience in the form of \emph{workflows}~\cite{wangagent,zhangaflow}, which explicitly capture stepwise task-solving processes, including intermediate reasoning steps, actions, and decision dependencies. By abstracting execution trajectories into node-structured processes~\cite{qiaobenchmarking}, workflow-based experience provides more structured and fine-grained guidance for subsequent task execution and adaptation. 

\textbf{Shared agent workflows.} To further improve workflow reuse and coordination at scale, recent work proposes sharing workflows across agents \cite{zhang2025g,zhangaflow}. In these settings, agents contribute structured workflows derived from execution trajectories to a shared repository \cite{tang2025agent}, which can be retrieved and adapted by others for new tasks. Shared workflows reduce redundant exploration and provide reusable procedural knowledge for long-horizon tasks \cite{qiaobenchmarking}.  Prior studies investigate workflows generated directly by language models \cite{wangagent} as well as workflows constructed through explicit structure exploration \cite{zhangaflow}, demonstrating improved efficiency and generalization in LLM-based agent systems. However, most approaches implicitly assume shared workflows are trustworthy and remain unaltered after publication, an assumption that becomes fragile in open or large-scale agent ecosystems \cite{srivastava2025memorygraft}.

\section{Threat Model}


We consider an agent ecosystem where agents upload execution trajectories to a cloud service. The service aggregates them into high-level workflows and maintains a shared repository. Downstream agents query this repository to retrieve related workflows for task execution.



\textbf{Adversary Setting.} We consider a threat setting in which adversarial behavior may arise from both centralized and decentralized sources within the system. 
The adversary attacks the system through two entry points in the system: \textit{(i) Centralized attack surface:} the shared repository, where stored trajectories and workflows can be modified after aggregation. \textit{(ii) Decentralized attack surface:} individual agent memory and trajectory generation process, where corrupted trajectories may be introduced before aggregation. Importantly, decentralized attacks are limited to a minority of agents, while the centralized repository constitutes the primary attack surface due to its system-wide influence. This assumption aligns with collaborative and federated learning, where participants are mostly honest and the aggregation service is modeled as the main point of compromise \cite{zhang2023survey,allouah2025towards,blanco2021achieving}. 

\textbf{Adversary Goal.} The adversary aims to construct a corrupted workflow that both bypasses the verification mechanism and induces erroneous execution in downstream agents. Formally, let $w$ denote a correct workflow derived from honest trajectories, and let $w'$ denote a modified workflow produced by the adversary. Let $\mathcal{V}(w') \in \{0,1\}$ denote the verification outcome, where $\mathcal{V}(w') = 1$ indicates that $w'$ is accepted by the verification procedure. Let $\mathcal{E}(w')$ denote the execution outcome of an agent following workflow $w'$, and let $\mathcal{E}(w)$ denote the correct execution outcome.

An attack is considered successful if and only if the following conditions hold:

\begin{equation}
\mathcal{A}(w') = 1 \iff \big(w' \neq w \ \wedge\ \mathcal{V}(w') = 1 \ \wedge\ \mathcal{E}(w') \neq \mathcal{E}(w)\big)
\end{equation}
Specifically, the adversary modifies the workflow ($w' \neq w$), the modified workflow is accepted by the verifier ($\mathcal{V}(w') = 1$), and execution leads to incorrect or unsafe outcomes ($\mathcal{E}(w') \neq \mathcal{E}(w)$).

\textbf{Adversary Capabilities.} We consider an adversary with capabilities spanning both centralized repository compromise and limited decentralized trajectory poisoning, as summarized in Table~\ref{tab:threat_model}. \textit{(C1) Repository-level compromise.} The adversary may fully compromise the centralized cloud repository, which serves as the primary attack surface, obtaining complete read and write access to stored trajectories and workflows. Under this capability, it can arbitrarily manipulate the shared repository layer, including deletion, insertion, modification, or reordering of workflow steps and stored trajectory records. The detailed attack procedures are included in the Appendix~\ref{attack_procedure}. \textit{(C2) Source agent-level poisoning.} In addition, the adversary may control a small subset of agents or their local memories, corresponding to a secondary attack surface at the trajectory submission stage, and inject corrupted or poisoned trajectories before they are submitted.

\textit{Security boundary.} We assume that cryptographic signatures on trajectories are unforgeable under standard assumptions, so that the adversary cannot produce validly signed but tampered trajectories. In addition, the adversary cannot compromise downstream agent runtimes or verification procedures, and cannot modify workflows or trajectories after retrieval. Importantly, the adversary’s influence is bounded: it may corrupt only a minority fraction of agents and their submitted trajectories, while the majority of trajectories associated with any workflow remain generated by honest agents, due to the inherently distributed nature of trajectory generation. \cite{zheng2026rethinking}

\begin{table}[t]
\vspace{-4pt}
\centering
\small
\setlength{\tabcolsep}{2.9pt}
\renewcommand{\arraystretch}{0.95}

\begin{tabularx}{\linewidth}{llX}
\toprule
\textbf{Category} & \textbf{Item} & \textbf{Specification} \\
\midrule

\multirow{2}{*}{Adversary Identity} 
& Location & Cloud-side shared repository + partial source agents \\
& Scope & Repository and a small subset of trajectory sources \\

\midrule

\multirow{2}{*}{Attack Surface} 
& Centralized repository & Modify stored workflows / trajectories \\
& Decentralized agent & Inject poisoned trajectories before submission \\

\midrule

\multirow{4}{*}{\makecell{Adversary \\ Capabilities}} 
& Repository Compromise & Full read/write access to centralized repository \\
& Workflow Manipulation & Deletion / insertion / modification / reordering \\
& Trajectory Manipulation & Modify stored trajectories (invalidates signatures) \\
& Source-level Poisoning & Control a minority of agents submitting corrupted trajectories\\

\midrule

\multirow{4}{*}{\makecell{Adversary\\ Limitations} }
& No Cryptographic Break & Cannot forge signatures or access signing keys \\
& Limited Compromise & Only a minority of agents can be corrupted     
\\
& No Execution-time Attack & No tampering during or after retrieval \\
& Scope Restriction & Restricted to repository and a minority of agents \\

\bottomrule
\end{tabularx}
\vspace{4pt}
\caption{Threat Model Specification: Adversary Identity, Capabilities, and Limitations}
\label{tab:threat_model}
\vspace{-16pt}
\end{table}



\section{Workflow Invalidation Detection}

This section presents our detection framework for defending against workflow invalidation attacks. We first introduce a trajectory-based integrity verification mechanism to filter corrupted trajectories and detect invalid workflows, followed by a runtime path-aware retrieval strategy to improve efficiency.

\subsection{Trajectory Provenance and Extraction} 

Each trajectory $\tau$ is represented as a DAG $G_\tau = (V_\tau, E_\tau)$, capturing a concrete execution instance. $V_\tau$ denotes the set of executed operations and $E_\tau$ encodes their realized dependency relations. Each operation in $V_\tau$ is aligned to an abstract workflow step. Importantly, this alignment can be \emph{many-to-one}, allowing multiple concrete operations to be abstracted into a single workflow step.

\paragraph{Traceability and Integrity Protection.} Before submission, each trajectory $\tau$ is canonically serialized as $\mathcal{C}(\tau)$ and signed using the originating agent’s private key: $
\sigma_\tau = \mathrm{Sign}_{sk}(H(\mathcal{C}(\tau)))
$. The signed tuple $(\tau, \sigma_\tau)$ is uploaded to the shared repository. Each trajectory is associated with its signer to maintain agent-level provenance. Only trajectories with valid signatures are admitted, and signature verification is enforced upon retrieval.

\paragraph{Trajectory Filtering.} Given a set of trajectories $\{\tau_i\}$ for the same task, we first perform trajectory-level cross-validation using multiple reasoning paths and structurally similar execution traces. Trajectories with inconsistent outcomes are flagged as anomalous. Second, for structurally similar trajectories, we encode each trajectory into an embedding $\mathbf{z}_i = f_{\mathrm{emb}}(\tau_i)$ and identify outliers based on embedding similarity. Trajectories are traced to their originating agents, and agents with repeated anomalous submissions are flagged. Their subsequent trajectories are subject to down-weighting during aggregation. Only trajectories passing validation are retained:
$
\mathcal{T}_{\mathrm{valid}} = \left\{ \tau_i \mid \tau_i \text{ passes validation} \right\}
$,
and used for workflow abstraction as well as evidence in downstream verification.

\subsection{Trajectory-Supported Workflow Verification}
\label{sec:verification}
We also model a workflow $w$ as a directed acyclic graph (DAG) $G_w = (V_w, E_w, \Phi)$, where $V_w$ is a set of workflow steps, $E_W \subseteq V_w \times V_w$ is a set of directed edges encoding execution dependencies, and $\Phi$ denotes step descriptions and branch conditions. Each node $v \in V_w$ represents an abstract operation (e.g., tool invocation, subtask, or decision step). Each edge $(u,v) \in E_w$ indicates that step $v$ may be executed only after step $u$ completes. Nodes with multiple outgoing edges denote branching points, where subsequent steps are selected based on execution context and conditions in $\Phi$, enabling the workflow to compactly represent a family of valid execution paths.

\textbf{Step-level alignment.} Our core design principle is to ground abstract workflows in verifiable execution evidence. Instead of trusting workflows as standalone artifacts, we require each workflow to be accompanied by a set of cryptographically signed execution trajectories, contributed by honest agents. These trajectories act as immutable provenance records, enabling downstream agents to validate both the integrity and the semantic faithfulness of retrieved workflows.

When a workflow $w$ is retrieved, the agent simultaneously requests a small set of cryptographically signed execution trajectories $\{(\tau_i, \sigma_{\tau_i})\}$ associated with $w$, i.e., trajectories used to extract or aligned with $w$ during workflow construction. The agent then performs \emph{cryptographic verification} , ensuring each signature is valid and the corresponding trajectory has not been modified after submission.





\textbf{Semantic consistency verification.} To validate workflow correctness beyond cryptographic integrity, the agent performs semantic consistency verification using supporting execution trajectories. Given a retrieved workflow $G_w = (V_w, E_w, \Phi)$ and a verified execution trajectory $G_\tau = (V_\tau, E_\tau)$, each trajectory operation is associated with a corresponding abstract workflow step. The verifier then checks whether the step descriptions, operation semantics, and dependency relations specified in $G_w$ are consistent with those realized in the authenticated execution evidence.

For a workflow step $v \in V_w$ that an agent is about to execute, the verifier requires the existence of the cryptographically verified execution trajectory $G_{\tau}$ whose corresponding subgraph provides concrete execution evidence for $v$. This subgraph consists of the set of trajectory operations that are aligned with the workflow step being verified. Specifically, the verifier checks the operations mapped to $v$ in $G_{\tau}$ are consistent with the intended functionality of $v$, and that the realized dependency relations in $E_\tau$ are compatible with the dependency constraints specified by $E_w$ and the branch conditions in $\Phi$.

We use a guided language model (GLM) $\mathcal{M}(\cdot)$ as the verifier. Given a retrieved workflow $G_w$ and a set of signed trajectories, it instantiates the predicate consistency through three complementary checks: (i). \textit{Structural consistency.} We enforce deterministic constraints on dependency relations. For each edge $(u, v) \in E_w$, the verifier checks whether a corresponding execution order or reachable path exists in $E_{\tau_i}$. (ii). \textit{Execution-level consistency.} the verifier retrieves aligned subgraphs from trajectories that provide concrete execution evidence, including operation types, inputs, and outputs. We then assess whether the state transitions induced by $v$ are consistent with those observed in the trajectories. (iii).  \textit{Semantic consistency.} Given the workflow specification, aligned trajectory subgraphs, and runtime context, the verifier produces a semantic consistency signal. 

\textbf{Multi-trajectory consensus.} 
Instead of relying on a single trajectory, we estimate the predicate consistency across multiple trajectories and express the fraction of trajectories supporting step \(v\) as
\begin{equation}
\alpha = \frac{1}{|\{\tau_i\}|} \sum_i \mathbf{1}\!\left[\text{Consistent}(v,\tau_i)\right]
\end{equation}
where \(\mathbf{1}[\cdot]\) is an indicator function equal to 1 if step \(v\) is consistent with trajectory \(\tau_i\), and 0 otherwise.

We accept a step $v$ only if (i) all deterministic constraints (both structural and execution-level) are satisfied, and (ii) $\alpha > \alpha_0$, where $\alpha_0$ is the minimum trajectory proportion consistent with the workflow. A workflow $w$ is accepted only if all its steps meet the above criteria. 

\subsection{Runtime Path-Aware Trajectory Retrieval} \label{sec:retrieval}
\label{subsec:pathaware}

\textbf{Motivation}. Workflows are inherently branching, and each execution trajectory provides evidence for only a subset of steps. Exhaustively retrieving all signed trajectories to cover the full workflow incurs prohibitive communication and verification overhead. We therefore propose a \emph{runtime path-aware trajectory retrieval strategy} that selectively retrieves trajectories relevant to the current execution context. Further algorithmic details are provided in Appendix~\ref{algorithm} .


\begin{figure}[t]
\centering
  \includegraphics[width=1\linewidth]{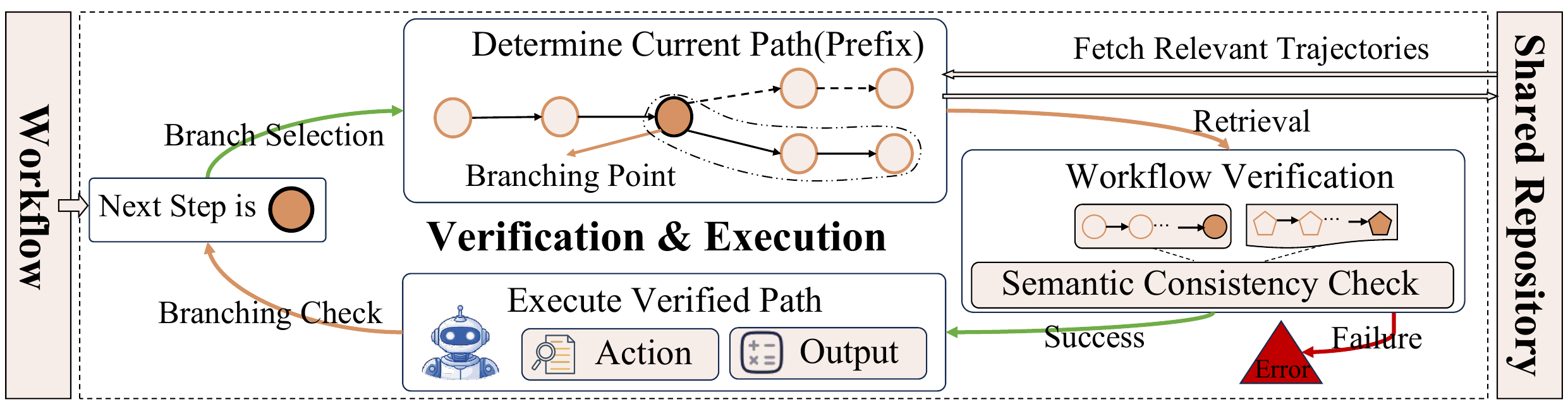}
  \vspace{-15pt}
  \caption{The agent retrieves a workflow and iteratively executes and verifies steps, checking them against supporting trajectories from the shared repository.}
  \vspace{-10pt}
  \label{path}

\end{figure}

\textbf{Path-aware retrieval scheme}.

In the path-aware trajectory retrieval scheme, supporting trajectories are retrieved and verified \emph{on demand} and restricted to the \emph{currently determined workflow path}. Here, the determined path refers to the sequence of workflow steps starting from the current execution point and extending up to the next branching node, for which the execution order is uniquely determined. As shown in Figure~\ref{path}, when a workflow is first retrieved, the agent fetches only the signed trajectories that support the initially determined path, i.e., the steps preceding the first branching point. As execution progresses, whenever the agent reaches a branching node, it selects the next branch based on the current execution context and the corresponding branch conditions, and subsequently requests only those signed trajectories whose aligned subgraphs are consistent with the steps along the selected branch. Execution then continues along the verified path until either all steps are completed or a branching point is reached. If at any point verification fails, the workflow is detected as invalid and execution is terminated. This design ensures that the agent continuously maintains authenticated execution evidence for workflow steps whose control flow has been determined, while avoiding unnecessary retrieval and verification of trajectories associated with non-executed branches.

\section{Evaluation}
\subsection{Setup}

\textbf{Models and Datasets.} We instantiate agents using different LLMs as the reasoning backbone, including LongChat-13B~\cite{longchat2023}, LLaMA3-8B~\cite{grattafiori2024llama}, and Qwen2.5-7B~\cite{bai2025qwen2}. To ensure a controlled comparison, the execution setup of the agent is kept fixed across all experiments. Unless otherwise specified, we deploy a local model (Qwen3-8B\cite{yang2025qwen3}) as the guided language model (GLM) for workflow invalidation detection.
We evaluate our method on four representative multi-step reasoning benchmarks: MathQA~\cite{amini2019mathqa}, MATH~\cite{hendrycks2021measuring}, MBPP~\cite{austin2021program}, and FEVER~\cite{thorne2018fever}. These benchmarks span diverse reasoning tasks, including mathematical problem solving (MathQA, MATH), program synthesis (MBPP), and multi-hop factual verification (FEVER). For each dataset, we randomly select 1K samples to construct execution trajectories, from which workflows are derived following the procedure in~\cite{zhangaflow,qiaobenchmarking}. To further evaluate generalization, we construct an additional synthetic dataset, denoted as \textit{Mix.}, by combining samples from all four benchmarks. 

\textbf{Baselines.} We compare our method against the following three workflow invalidation detection baselines: 
\textit{(i) Semantic evaluation (Sem\_eva).} This baseline follows the LLM-as-a-judge paradigm and evaluates workflows by prompting the model to assess the semantic correctness and coherence of intermediate reasoning steps. It detects workflow invalidation by identifying semantic inconsistencies during execution, motivated by~\cite{zhou2025trustrag}. 
\textit{(ii) Consistency comparison (Con\_com).} This baseline adopts a consistency-based detection strategy, following the spirit of~\cite{wei2025memguard}. It retrieves the top-$k$ candidate workflows and identifies those that lead to inconsistent reasoning outcomes as invalid. 
\textit{(iii) Embedding evaluation (Emb\_eva).} This baseline measures embedding-level semantic representations of workflows. Candidate workflows are encoded into a shared embedding space, and workflows exhibiting low agreement relative to others are identified as invalid, motivated by~\cite{besta2024checkembed}.

\textbf{Metrics.} To quantify the effectiveness of workflow invalidation attacks, we compare inference accuracy on the dataset with and without workflow invalidation attacks. The resulting accuracy drop directly reflects the degradation in inference quality caused by the workflow invalidation attack. In addition, we report the \textit{Attack Success Rate (ASR)} (\%), defined as the fraction of inference samples for which a previously correct execution becomes incorrect due to workflow invalidation. Specifically, let $N_{correct}$ denote the number of tasks correctly solved under the original workflows. After applying workflow invalidation, $N_{acc}$ denotes the number of tasks that remain correct. The ASR is defined as: $\mathrm{ASR} = (1 - \frac{N_{\text{acc}}}{N_{\text{correct}}})\times 100\%$. Note that ASR is computed over the subset of samples that are correct under the original workflow, and thus is not equivalent to the absolute accuracy drop.

To evaluate detection performance, we report \textit{True Positive Rate (TPR)} (\%), \textit{True Negative Rate (TNR)} (\%), and \textit{F1 score} (\%). TPR denotes correctly detected invalid workflows, while TNR measures correctly identified valid ones. The F1 score captures overall robustness via precision-recall trade-off.


We additionally provide detailed experimental specifics in Appendix~\ref{setup_verification}, including prompt design and example outputs.We also report the robustness of trajectory selection in Appendix~\ref{app:trstudy}, and time complexity analysis and detailed detection overhead results in Appendix~\ref{app:ablation_study}.

\begin{table*}[h]
\vspace{-4pt}
\renewcommand{\arraystretch}{1.2}
\centering
\resizebox{\linewidth}{!}{
\begin{tabular}{c|c|c|cc|cc|cc|cc}
\hline
\multirow{2}{*}{\textbf{Agent LLM}} 
& \multirow{2}{*}{\textbf{Dataset}} 
& \textbf{Base} 
& \multicolumn{2}{c|}{\textbf{Deletion}} 
& \multicolumn{2}{c|}{\textbf{Insertion}} 
& \multicolumn{2}{c|}{\textbf{Modification}} 
& \multicolumn{2}{c}{\textbf{Reordering}} \\
\cline{3-11}
& 
& ACC& ACC& ASR& ACC& ASR& ACC& ASR& ACC& ASR\\
\hline

\multirow{5}{*}{\textbf{LongChat-13B}}& MathQA 
& 83.53 
& 15.44$_{\textcolor{red}{\downarrow 68.09}}$ & 81.52 
& 21.79$_{\textcolor{red}{\downarrow 61.74}}$ & 73.91 
& 35.64$_{\textcolor{red}{\downarrow 47.89}}$ & 57.33 
& 10.37$_{\textcolor{red}{\downarrow 73.16}}$ & 87.59 \\

& MATH 
& 88.82 
& 28.74$_{\textcolor{red}{\downarrow 60.08}}$ & 67.64 
& 18.17$_{\textcolor{red}{\downarrow 70.65}}$ & 79.54 
& 33.53$_{\textcolor{red}{\downarrow 55.29}}$ & 62.25 
& 16.35$_{\textcolor{red}{\downarrow 72.47}}$ & 81.59 \\

& FEVER 
& 61.07 
& 12.95$_{\textcolor{red}{\downarrow 48.12}}$ & 88.36 
& 13.62$_{\textcolor{red}{\downarrow 47.45}}$ & 73.91 
& 22.48$_{\textcolor{red}{\downarrow 38.59}}$ & 69.81 
& 29.27$_{\textcolor{red}{\downarrow 31.80}}$ & 64.93 \\

& MBPP 
& 76.89 
& 21.56$_{\textcolor{red}{\downarrow 55.33}}$ & 82.04 
& 35.42$_{\textcolor{red}{\downarrow 41.47}}$ & 59.05 
& 20.08$_{\textcolor{red}{\downarrow 56.81}}$ & 73.12 
& 11.43$_{\textcolor{red}{\downarrow 65.46}}$ & 79.87 \\

& Mix. 
& 77.58 
& 19.67$_{\textcolor{red}{\downarrow 57.91}}$ & 75.35 
& 22.25$_{\textcolor{red}{\downarrow 55.33}}$ & 68.69 
& 27.93$_{\textcolor{red}{\downarrow 49.65}}$ & 67.03 
& 16.86$_{\textcolor{red}{\downarrow 60.72}}$ & 76.42 \\

\hline

\multirow{5}{*}{\textbf{LLaMA3.1-8B}} 
& MathQA 
& 68.57 
& 12.73$_{\textcolor{red}{\downarrow 55.84}}$ & 80.19& 11.43$_{\textcolor{red}{\downarrow 57.14}}$ & 83.33& 25.56$_{\textcolor{red}{\downarrow 43.01}}$ & 63.02& 12.30$_{\textcolor{red}{\downarrow 56.27}}$ & 82.87\\

& MATH 
& 42.73 
& 8.33$_{\textcolor{red}{\downarrow 34.40}}$ & 80.05& 9.17$_{\textcolor{red}{\downarrow 33.56}}$ & 84.25& 3.25$_{\textcolor{red}{\downarrow 39.48}}$ & 92.44& 11.48$_{\textcolor{red}{\downarrow 31.25}}$ & 74.31\\

& FEVER 
& 34.21 
& 4.36$_{\textcolor{red}{\downarrow 29.85}}$ & 87.28& 10.88$_{\textcolor{red}{\downarrow 23.33}}$& 70.15& 13.01$_{\textcolor{red}{\downarrow 21.20}}$ & 65.30& 16.94$_{\textcolor{red}{\downarrow 17.27}}$ & 50.54\\

& MBPP 
& 56.15 
& 17.94$_{\textcolor{red}{\downarrow 38.21}}$ & 76.58& 18.09$_{\textcolor{red}{\downarrow 38.06}}$ & 68.72& 9.54$_{\textcolor{red}{\downarrow 46.61}}$ & 83.07& 15.28$_{\textcolor{red}{\downarrow 40.87}}$ & 73.18\\

& Mix. 
& 50.42 
& 10.84$_{\textcolor{red}{\downarrow 39.58}}$& 77.32& 12.39$_{\textcolor{red}{\downarrow 38.03}}$ & 75.46& 12.84$_{\textcolor{red}{\downarrow 37.58}}$ & 74.35& 14.00$_{\textcolor{red}{\downarrow 36.42}}$ & 70.22\\

\hline

\multirow{5}{*}{\textbf{Qwen2.5-7B}}& MathQA & 68.93& 21.40$_{\textcolor{red}{\downarrow 47.53}}$& 71.21 & 18.57$_{\textcolor{red}{\downarrow 50.36}}$& 74.62 & 10.71$_{\textcolor{red}{\downarrow 58.22}}$& 77.83 & 19.65$_{\textcolor{red}{\downarrow 49.28}}$& 70.68 \\
& MATH   & 60.42& 13.97$_{\textcolor{red}{\downarrow 46.45}}$& 80.09 & 20.01$_{\textcolor{red}{\downarrow 40.41}}$& 68.67 & 16.67$_{\textcolor{red}{\downarrow 43.75}}$& 75.43 & 13.28$_{\textcolor{red}{\downarrow 47.14}}$& 81.28 \\
& FEVER  & 48.03& 13.26$_{\textcolor{red}{\downarrow 34.77}}$& 79.02& 10.37$_{\textcolor{red}{\downarrow 37.66}}$& 85.59& 19.41$_{\textcolor{red}{\downarrow 28.62}}$& 65.05& 8.99$_{\textcolor{red}{\downarrow 39.04}}$& 88.73\\
& MBPP   & 57.35& 7.19$_{\textcolor{red}{\downarrow 50.16}}$& 90.12& 14.62$_{\textcolor{red}{\downarrow 42.73}}$& 77.69& 28.57$_{\textcolor{red}{\downarrow 28.78}}$& 52.33& 21.05$_{\textcolor{red}{\downarrow 36.30}}$& 64.82\\
& Mix.   & 58.68& 13.96$_{\textcolor{red}{\downarrow 44.72}}$& 80.11& 15.89$_{\textcolor{red}{\downarrow 47.79}}$& 76.64& 18.84$_{\textcolor{red}{\downarrow 39.84}}$& 67.66& 16.56$_{\textcolor{red}{\downarrow 42.12}}$& 76.38\\

\hline
\end{tabular}
}
\caption{Effectiveness of the proposed workflow invalidation attack.}
\label{tab:attraction_results}
\vspace{-10pt}
\end{table*}

\subsection{Overall Performance}

\textbf{Effectiveness of workflow invalidation.} We evaluate the effectiveness of workflow invalidation attacks by analyzing how adversarial workflow modifications affect agent task execution. Experiments are conducted on four representative multi-step reasoning datasets. For each dataset, an adversary applies a range of workflow modifications, including deletion of critical steps, insertion of misleading steps, substitution of key steps, and reordering of dependent steps. Attack effectiveness is measured using inference accuracy and ASR under correct and invalidated workflows.
As shown in Table~\ref{tab:attraction_results}, agents achieve consistently high inference accuracy when guided by correct workflows across all benchmarks. In contrast, workflow invalidation leads to severe performance degradation, which can reduce agent inference accuracy by up to 73.16 percentage points, with the average accuracy dropping to 16.83\% under invalid workflows. Among the considered attack strategies, step substitution and insertion are the most damaging, achieving an average ASR of 75.60\%. Step deletion and reordering also substantially degrade performance, demonstrating that adversarially invalidated workflows can reliably mislead agent execution across diverse reasoning tasks. Specifically, for datasets where agents already achieve relatively moderate performance under correct workflows, such as FEVER, adversarial modifications still lead to noticeable performance degradation, indicating that invalid workflows can consistently harm execution even in challenging settings. In contrast, for datasets where agents perform well under correct workflows, such as MathQA, workflow invalidation results in more pronounced performance drops and higher attack success rates. It suggests that tasks which are more sensitive to workflow guidance are more vulnerable to adversarial workflow manipulations.

\begin{table*}[h]
\centering
\setlength{\tabcolsep}{3.1pt} 
\scriptsize
\resizebox{\linewidth}{!}{
\begin{tabular}{c|ccc|ccc|ccc|ccc|ccc}
\toprule
\multirow{2}{*}{\textbf{Method}}
& \multicolumn{3}{c|}{\textbf{MathQA}}
& \multicolumn{3}{c|}{\textbf{MATH}}
& \multicolumn{3}{c|}{\textbf{FEVER}}
& \multicolumn{3}{c|}{\textbf{MBPP}}
& \multicolumn{3}{c}{\textbf{Mix}} \\
\cmidrule(lr){2-4} \cmidrule(lr){5-7} \cmidrule(lr){8-10} \cmidrule(lr){11-13} \cmidrule(lr){14-16}
& TPR & TNR & F1
& TPR & TNR & F1
& TPR & TNR & F1
& TPR & TNR & F1
& TPR & TNR & F1 \\
\midrule
Sem\_eva & 57.28 & 74.95 & 65.93
& 48.32 & 58.01 & 58.05
& 46.17 & 53.90 & 57.87
& 34.22 & 68.20 & 42.95
& 49.20& 62.79 & 58.04\\
Con\_com & 52.18 & 23.51 & 53.67
& 29.73 & 56.26 & 40.08
& 21.15 & 72.48 & 30.52
& 43.84 & 22.73 & 25.20
& 45.36 & 49.00 & 51.49 \\
Emb\_eva & 44.44 & 25.45 & 48.76
& 34.29 & 31.86 & 32.06
& 21.95 & 39.75 & 32.09& 26.17 & 49.72 & 32.52
& 39.21 & 31.84 & 44.02 \\
\rowcolor{gray!20}
\textbf{Ours*} & \textbf{82.27} & \textbf{91.58} & \textbf{87.88}
& \textbf{78.10} & \textbf{89.51} & \textbf{85.56}
& \textbf{80.64} & \textbf{76.59} & \textbf{82.59}
& \textbf{78.33} & \textbf{83.94} & \textbf{83.85}
& \textbf{80.51} & \textbf{85.41} & \textbf{85.26} \\
\bottomrule
\end{tabular}
}
\vspace{-2pt}
\caption{Comparison of different workflow invalidation detection methods, including the Mix dataset.}
\label{tab:detection_results}
\vspace{-5pt}
\end{table*}

\textbf{Detection performance.} Table~\ref{tab:detection_results} presents workflow invalidation detection performance across four datasets. Our method consistently outperforms all baselines in TPR, TNR, and F1 score, demonstrating superior detection of invalid workflows while preserving valid ones.

The baseline methods exhibit distinct weaknesses. Sem\_eva achieves moderate TPR but relatively low F1 scores, as in some datasets, the method degenerates into over-predicting invalid samples, resulting in imbalanced performance where only one of TPR or TNR remains high. This suggests that semantic-based judgments are insufficient for reliably distinguishing workflows. Con\_com shows highly imbalanced behavior, with either low TPR or low TNR depending on the dataset. This suggests that outcome-consistency signals are inherently unstable for detection, since workflow invalidation does not necessarily manifest as inconsistent execution outcomes, and the method further deteriorates as the proportion of invalid workflows in the dataset increases. Emb\_eva performs the worst overall, reflecting the limited sensitivity of embedding-level agreement to fine-grained semantic and structural violations in workflows.

In contrast, our method achieves both high TPR and high TNR across all datasets, resulting in substantially higher F1 scores. On average, our approach improves F1 by more than 40.14 percentage points over baselines, indicating a better capability for workflow invalidation detection. Notably, our method achieves an average TPR of 79.97\% across all datasets, highlighting its effectiveness in detecting invalid workflows in complex, multi-step reasoning scenarios. The gains observed on FEVER further demonstrate the robustness of our methods across diverse task domains. This confirms that our approach provides a reliable and generalizable solution across heterogeneous benchmarks. 

\begin{figure*}[ht]
    \centering
    \includegraphics[width=\textwidth]{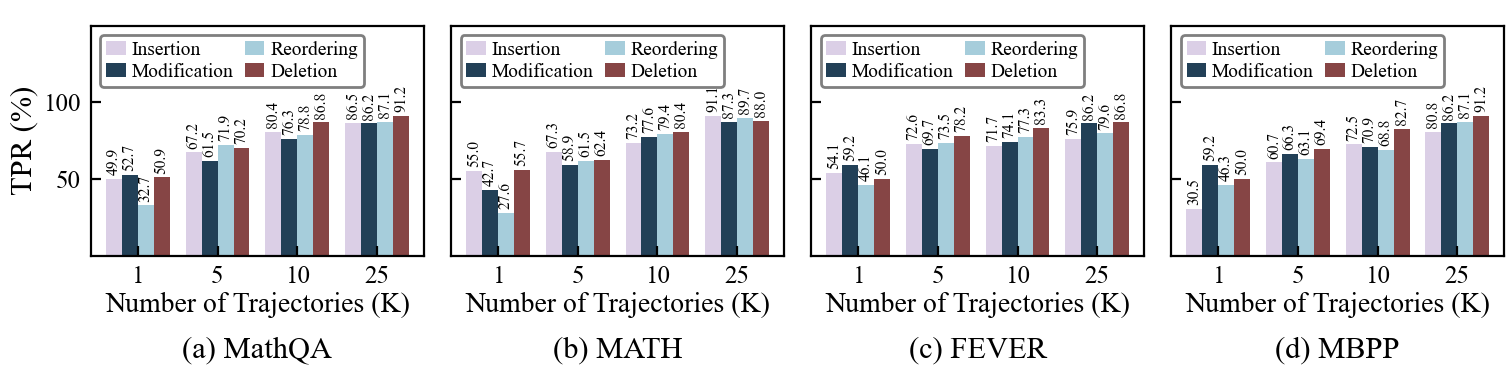}
      \vspace{-10pt}
    \caption{Performance of the workflow invalidation detection with different number of trajectories.}

    \label{fig:abl_k_traj}

\end{figure*}

\subsection{Ablation Study}

\textbf{Impact of the number of trajectories for detection.} We further analyze how the number of supporting trajectories affects workflow invalidation detection performance. Specifically, we vary the number of trajectories $K$ from 1 to 25 and report the invalid workflow detection rate (TPR) under different workflow invalidation attacks. Figure~\ref{fig:abl_k_traj} summarizes the results across four benchmarks. Overall, we observe a consistent trend across all datasets: detection performance improves substantially as $K$ increases from 1 to 10, and then gradually saturates when $K>10$. This behavior highlights the effectiveness of aggregating execution evidence from multiple trajectories: when only a single trajectory is available, detection accuracy is highly sensitive to randomness and noise in individual executions, especially under subtle attacks such as \textit{Insertion} and \textit{Reordering}. As $K$ increases, aggregating multiple trajectories suppresses the influence of outlier executions and enables the verifier to better capture stable execution patterns, resulting in significant improvements across all attack types. 

We also observe notable differences across different invalidation attacks. In general, \textit{Deletion} and \textit{Modification} attacks are detected more reliably with fewer trajectories, as they tend to introduce more pronounced inconsistencies between workflows and execution trajectories. In contrast, \textit{Insertion} and \textit{Reordering} attacks require a larger number of trajectories to reach comparable detection accuracy, since these attacks are designed to preserve local plausibility while subtly altering execution order or intermediate states. 
Nevertheless, once $K$ reaches around 10, detection performance for all attack types converges to a high and stable level. Beyond $K=10$, additional trajectories provide diminishing marginal benefits. This is because newly trajectories are often highly similar to existing ones and contribute limited new structural or semantic information for verification.

\textbf{Impact of different GLMs for detection.} We next evaluate the impact of different guided language models (GLMs) used for workflow invalidation detection (\S\ref{sec:verification}). Specifically, we consider three representative models with different scales and capacities: Qwen3-8B \cite{yang2025qwen3}, LLaMA3-8B \cite{grattafiori2024llama}, and Qwen2.5-7B \cite{bai2025qwen2}. The results are summarized in Table~\ref{tab:model_results}. Overall, all three GLMs achieve strong detection performance across datasets. In particular, the TPR consistently remains around 75.39\% across all settings, while the average TNR stays around 81.52\%. This indicates that the proposed detection method generalizes well across different underlying GLMs, as workflow invalidation detection is largely driven by the verification procedure and execution trajectories, rather than by any specific model architecture or scale.

We observe that larger GLMs, such as Qwen3-8B, achieve more balanced TPR-TNR trade-offs and consistently deliver the highest F1 scores across datasets, with particularly strong gains on complex reasoning benchmarks such as MathQA and MATH that require robust semantic understanding and long-horizon reasoning. LLaMA3-8B shows comparable performance, indicating that large models across architectures can support workflow verification. In contrast, the smaller Qwen2.5-7B exhibits lower F1 scores, especially on FEVER and MBPP, where multi-hop factual reasoning and program-level semantics pose greater challenges. Nevertheless, the performance degradation is limited, and the model maintains competitive TPR and TNR values. This reflects a favorable accuracy-efficiency trade-off, enabling lightweight GLMs suitable under resource or latency constraints.

\begin{table*}[t]
\centering
\setlength{\tabcolsep}{3.1pt} 
\scriptsize 
\resizebox{\linewidth}{!}{
\small
\begin{tabular}{c| ccc| ccc| ccc| ccc| ccc}
\toprule
\multirow{2}{*}{\textbf{GLM}}
& \multicolumn{3}{c|}{\textbf{MathQA}} 
& \multicolumn{3}{c|}{\textbf{MATH}} 
& \multicolumn{3}{c}{\textbf{FEVER}}
& \multicolumn{3}{c}{\textbf{MBPP}}
& \multicolumn{3}{c}{\textbf{Mix}}\\
\cmidrule(lr){2-4} \cmidrule(lr){5-7} \cmidrule(lr){8-10} \cmidrule(lr){11-13} \cmidrule(lr){14-16}
& TPR & TNR& F1& TPR & TNR& F1& TPR & TNR& F1& TPR & TNR& F1& TPR & TNR& F1\\
\midrule
\textbf{Qwen3-8B}&  80.64&  91.58&  87.88& 77.10&  89.51&  85.56& 76.91&  76.59&  82.59& 74.33& 83.94& 83.85& 78.50&  85.41&  85.26\\
\textbf{LLaMA3-8B}&  76.92&  85.71&  83.91& 76.53&  85.84&  80.75&  73.44&  71.02&  73.99&  73.98& 78.59& 76.05& 74.68&  79.38&  78.74\\
\textbf{Qwen2.5-7B}&  75.96&  87.08&  79.45& 76.93&  81.82&  78.64&  68.91&  67.39&  68.03& 72.13& 79.89& 75.39&  73.87&  79.24&  76.18\\
\bottomrule
\end{tabular}
}
\caption{The performance of workflow invalidation detection with different guided language models.}
\label{tab:model_results}
\vspace{-12pt}
\end{table*}

\section{Conclusion}
In this work, we identify workflow invalidation as a new and critical attack surface in LLM-based agent systems that rely on shared workflows. We show that adversarially tampered workflows can substantially degrade agent performance. To defend against this attack, we propose a workflow detection framework based on cryptographically signed execution trajectories and runtime verification, enabling agents to detect invalidated workflows during execution. Our path-aware trajectory retrieval strategy further ensures that verification remains efficient and scalable by restricting trajectory retrieval to the currently executed workflow path. Extensive experiments across multiple datasets and model configurations demonstrate both the severity of workflow invalidation attacks and the effectiveness of our defense. We also discuss current limitations and future extensions to broader adversarial settings in Appendix~\ref{app:lf}

\section{Impact Statements}
This paper presents work whose goal is to advance the field of machine learning. There are many potential societal consequences of our work, none of which we feel must be specifically highlighted here.

\bibliographystyle{unsrt}

\bibliography{refer}

\newpage
\appendix
\onecolumn

\section{Algorithm Design for Path-aware Workflow Verification}
\label{algorithm}

Algorithm~\ref{alg:pathaware} presents the runtime path-aware workflow verification procedure. The algorithm incrementally verifies and executes workflow steps by alternating between verification and execution phases, while dynamically adapting to branching workflow steps.

Given a workflow path $\mathcal{P}$ whose execution order has been determined, \textsc{Workflow\_Ver} verifies each step $u \in \mathcal{P}$ independently. For each step, the verifier retrieves a small number of signed execution trajectories aligned with $u$ and validates their signatures to ensure trajectory integrity. It then checks whether the workflow step is semantically consistent with the authenticated execution evidence derived from these trajectories. If any step fails verification, a workflow invalidation attack is detected and the procedure terminates immediately. Otherwise, all steps in $\mathcal{P}$ are accepted, and a verified path $\mathcal{P}^{*}$ is returned.

The verified path $\mathcal{P}^{}$ is then passed to \textsc{Agent\_Exe}, which executes all steps in $\mathcal{P}^{*}$ in the runtime environment. After executing the verified steps, the agent checks whether the next step corresponds to a branching node. If a branching node is encountered, the agent determines the next workflow path based on the current execution context and the corresponding branch conditions. The agent then returns the newly determined path $\mathcal{P}$ for subsequent verification. Otherwise, if no further branching occurs, the agent execution completes.

By verifying and executing only the currently determined workflow path, the algorithm performs verification on demand and confines trajectory retrieval to the execution path actually taken. This design minimizes verification overhead while preserving strong integrity and semantic consistency guarantees against workflow invalidation attacks.

\begin{algorithm}[H]
\caption{Runtime Path-Aware Workflow Verification}
\label{alg:pathaware}
\begin{algorithmic}
    \Procedure{\textsc{Workflow\_Ver}}{$\mathcal{P}$}
        \Comment{Verify the currently determined workflow path $\mathcal{P}$ step by step}
        \ForAll{step $u \in \mathcal{P}$}
            \State Fetch $K$ signed execution trajectories aligned with workflow step $u$, denoted as $\mathcal{R}_u$
            \State Verify the cryptographic signatures of trajectories in $\mathcal{R}_u$, obtaining $\mathcal{R}_u^{*}$
            \State Verify the semantic consistency of workflow step $u$ using $\mathcal{R}_u^{*}$
            \If{step $u$ does \textit{not} pass verification}
                \State \Return Workflow invalidation detected
            \EndIf
        \EndFor
        \State \Return Verified path $\mathcal{P}^{*}$
    \EndProcedure
    \\
    \Procedure{\textsc{Agent\_Exe}}{$\mathcal{P}^{*}$}
        \Comment{Execute all verified steps in the runtime environment}
        \State Execute all workflow steps in the verified path $\mathcal{P}^{*}$
        \If{the next step $v$ is a branching node}
            \State Determine the workflow path based on the current branching node
            \State $\mathcal{P} \gets \textsc{DeterminePath}(G_w, v)$
            \State \textsc{Workflow\_Ver}($\mathcal{P}$) \Comment{handoff to \textsc{Workflow\_Ver}}
        \Else
            \State \Return Agent execution completes
        \EndIf
    \EndProcedure
\end{algorithmic}
\end{algorithm}


\section{Workflow Invalidation Procedure}\label{attack_procedure}
\begin{figure}[t]

    \centerline{\includegraphics[width=0.9\textwidth]{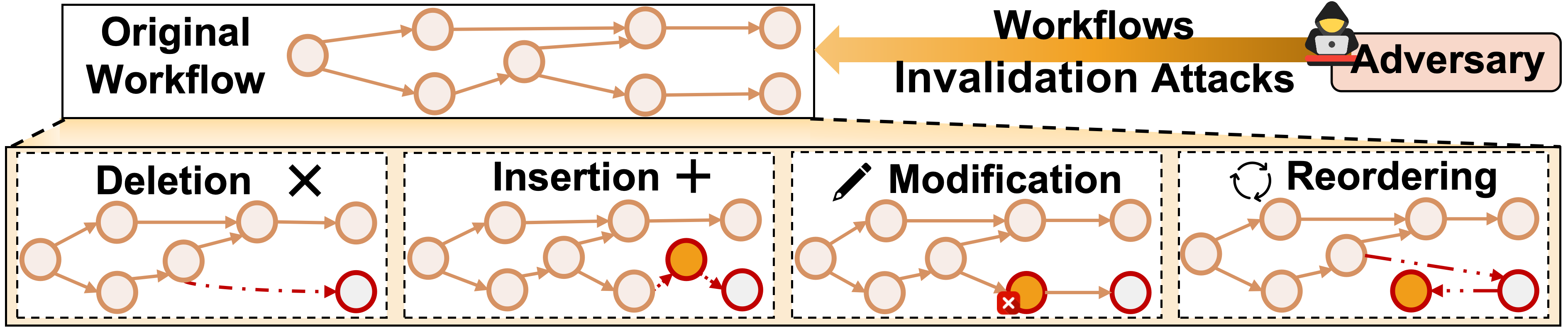}}
\caption{Different strategies for the workflow invalidation attack.}
\label{Threat Model}
\vspace{-6pt}
\end{figure}

The adversary does not introduce obvious syntactic errors or explicit illegal instructions, but instead invalidates the workflow by altering the execution path induced by the workflow. As shown in Figure~\ref{Threat Model}, the adversary can perform the following workflow modification strategies, each designed to preserve surface-level plausibility while misleading agents' execution. 
\begin{itemize}
    \item \textbf{Deletion:} The adversary removes critical steps essential for correct execution. Although seemingly minor in isolation, these steps maintain intermediate states and dependencies, and their removal causes subsequent steps to operate on incomplete or inconsistent states, ultimately affecting the intended outcome.
    \item \textbf{Insertion:} The adversary inserts additional steps into the workflow, such as extra validation, preprocessing, or refinement steps. These inserted steps interfere with the normal execution logic and propagate errors to subsequent steps, ultimately causing the workflow to deviate from its intended behavior.
    \item \textbf{Modification:} The adversary modifies the content of existing steps by altering conditions, parameters, or decision criteria, while preserving the overall intent of the workflow. These modifications maintain local semantic coherence, but shift execution logic in ways that accumulate into incorrect outcomes over multiple steps.
    \item \textbf{Reordering:} The adversary reorders dependent workflow steps in a manner that respects superficial logical flow, but violates implicit temporal or causal dependencies. Such reordering is difficult to detect through local inspection, yet can invalidate assumptions about intermediate state availability and execution order.
\end{itemize}
Collectively, these attacks do not rely on explicit semantic corruption or syntactic invalidity. Instead, they exploit the fact that workflow correctness depends on global structural consistency and long-range execution dependencies, which are not directly observable through step-level inspection.

\section{Detailed Experimental Setup for Workflow Verification}
\label{setup_verification}
This section describes the prompt design used in our runtime path-aware workflow verification experiments. The prompt specifies the verifier’s role, input assumptions, and consistency criteria, and defines a structured output format that reports both the consistency decision and the inferred type of workflow invalidation. 
\subsection{Prompt Design}
We design a path-aware verification prompt that consists of a \textit{task definition} specifying the verifier’s role and inputs, and a set of \textit{verification rules} that require the LLM to determine whether the executed prefix and upcoming step are supported by authenticated trajectories.
\begin{promptbox}{\Large\textbf{Prompt 1 (Task Definition)}}

\noindent\textbf{System:} You are an expert verifier for \textsc{One-Variable Linear-Equation} workflows.

\vspace{0.35em}
\noindent\textbf{Goal:} Verify whether a given workflow (or workflow step/module) is semantically consistent with and supported by a small set of signed execution trajectories. Detect workflow--trajectory mismatches that may indicate workflow invalidation attacks (e.g., deletion, insertion, substitution, or reordering), even when the workflow appears syntactically valid and superficially plausible.

\vspace{0.35em}
\noindent\textbf{Inputs:}
\begin{enumerate}[leftmargin=1.2em, itemsep=0.15em, topsep=0.15em]
  \item \textbf{Workflow:} A workflow candidate in JSON (or a single workflow step), including a declared module code. The workflow library contains:
  \begin{itemize}[leftmargin=1.2em, itemsep=0.15em, topsep=0.15em]
    \item \textbf{Level-1 (L1) atomic modules:}
    \begin{itemize}[leftmargin=1.2em, itemsep=0.1em, topsep=0.1em]
      \item \texttt{L1-LIN-ISOLATE}
      \item \texttt{L1-LIN-FRACTION-LCD}
      \item \texttt{L1-LIN-RECIPROCAL}
      \item \texttt{L1-LIN-DISTR-COLLECT}
      \item \texttt{L1-LIN-PROPORTION}
    \end{itemize}

    \item \textbf{Level-2 (L2) composite modules (ordered compositions of L1):}
    \begin{itemize}[leftmargin=1.2em, itemsep=0.1em, topsep=0.1em]
      \item \texttt{L2-LIN-FRAC-TO-ISOLATE} $=$ [\texttt{L1-LIN-FRACTION-LCD} $\rightarrow$ \texttt{L1-LIN-ISOLATE}]
      \item \texttt{L2-LIN-DISTR-TO-ISOLATE} $=$ [\texttt{L1-LIN-DISTR-COLLECT} $\rightarrow$ \texttt{L1-LIN-ISOLATE}]
      \item \texttt{L2-LIN-PROP-TO-ISOLATE} $=$ [\texttt{L1-LIN-PROPORTION} $\rightarrow$ \texttt{L1-LIN-ISOLATE}]
      \item \texttt{L2-LIN-RECIP-TO-ISOLATE} $=$ [\texttt{L1-LIN-RECIPROCAL} $\rightarrow$ \texttt{L1-LIN-ISOLATE}]
    \end{itemize}
  \end{itemize}

  \item \textbf{Trajectories:} A small collection of execution trajectories aligned with the workflow. Each trajectory contains:
  \begin{itemize}[leftmargin=1.2em, itemsep=0.15em, topsep=0.15em]
    \item a problem/query
    \item a response/execution text (step description)
  \end{itemize}
\end{enumerate}

\vspace{0.35em}

\noindent\textbf{Assumption:} Trajectories are assumed authentic (signatures already verified). Evaluate semantic consistency only.

\vspace{0.35em}
\noindent\textbf{Task:} For each workflow (or step), determine whether the declared workflow module and step description are correctly supported by the provided execution trajectories.

\vspace{0.35em}
\noindent\textbf{Important constraints:}
\begin{itemize}[leftmargin=1.2em, itemsep=0.2em, topsep=0.2em]
  \item Do not recompute final numeric answers.
  \item Only analyze the methodology / step sequence implied by the trajectories' responses.
  \item Treat trajectories as execution evidence: if the workflow claims a reasoning pattern not exhibited in trajectories, mark it as \texttt{NOT\_CONSISTENT}.
  \item If trajectories support multiple reasoning strategies, judge whether the declared module remains a reasonable dominant abstraction.
  \item If the item is not a one-variable linear-equation solver task, mark \texttt{NOT\_CONSISTENT} with reason ``not a one-variable linear-equation task''.
\end{itemize}
\end{promptbox}

\begin{promptbox}{\Large\textbf{Prompt 2 (Verification Rules)}}
\noindent\textbf{Consistency rules (L1):}
\begin{itemize}[leftmargin=1.2em, itemsep=0.15em, topsep=0.15em]
  \item If the declared module is an L1 atomic module, trajectories must clearly exhibit the decisive reasoning steps of that module.
  \item Minor omissions (e.g., not explicitly stating domain restrictions) do not invalidate consistency if the method is otherwise clear; note the omission briefly.
  \item If trajectories are too vague to infer a method, mark \texttt{NOT\_CONSISTENT} with reason ``insufficient evidence''.
\end{itemize}

\vspace{0.35em}
\noindent\textbf{Consistency rules (L2):}
\begin{itemize}[leftmargin=1.2em, itemsep=0.15em, topsep=0.15em]
  \item If the declared module is an L2 composite module, trajectories must provide evidence for each sub-stage in the correct order.
  \item A workflow is consistent with an L2 module only if decisive steps for component L1 modules are observed in trajectories in the declared order.
  \item If trajectories only support a subset of the L2 stages, mark \texttt{NOT\_CONSISTENT} and report the best supported alternative (L1 or L2).
\end{itemize}

\vspace{0.35em}
\noindent\textbf{Attack type annotation:}
\begin{itemize}[leftmargin=1.2em, itemsep=0.15em, topsep=0.15em]
  \item If a workflow is marked \texttt{NOT\_CONSISTENT}, infer the most likely invalidation pattern:
  \begin{itemize}[leftmargin=1.2em, itemsep=0.1em, topsep=0.1em]
    \item \texttt{deletion}: critical reasoning steps implied by trajectories are missing from the workflow
    \item \texttt{insertion}: workflow contains extra steps not supported by trajectories
    \item \texttt{substitution}: workflow claims a different reasoning strategy than trajectories exhibit
    \item \texttt{reordering}: workflow step order violates dependencies implied by trajectories
    \item \texttt{unclear}: inconsistency detected but attack pattern cannot be confidently determined
  \end{itemize}
\end{itemize}
\end{promptbox}

\subsection{Output Format}

The output consists of two components:
 (i). a binary consistency decision indicating whether the workflow execution is valid, and (ii). a categorical label identifying the inferred type of workflow invalidation, if any.

\begin{promptbox}{\Large\textbf{Output Format}}

\noindent\textbf{Return JSON only (no code fences, no extra prose) with schema:}

\small
\begin{verbatim}
{
  "summary": {
    "total_workflows": <int>,
    "consistent_count": <int>,
    "inconsistent_count": <int>,
    "by_module": {
      "L1-LIN-ISOLATE": <int>,
      "L1-LIN-FRACTION-LCD": <int>,
      "L1-LIN-RECIPROCAL": <int>,
      "L1-LIN-DISTR-COLLECT": <int>,
      "L1-LIN-PROPORTION": <int>,
      "L2-LIN-FRAC-TO-ISOLATE": <int>,
      "L2-LIN-DISTR-TO-ISOLATE": <int>,
      "L2-LIN-PROP-TO-ISOLATE": <int>,
      "L2-LIN-RECIP-TO-ISOLATE": <int>
    }
  },
  "workflows": [
    {
      "index": <1-based index>,
      "declared_module": "...",
      "workflow_snippet": "<short snippet>",
      "consistent": true/false,
      "supported_module": "... | null",
      "supported_decomposition": ["<module1>", "<module2>"] | null,
      "suspected_attack_type": 
      "deletion | insertion | substitution | reordering | unclear | null",
      "reason": "<brief explanation>"
    }
  ]
}
\end{verbatim}
\normalsize
\end{promptbox}

\section{The Robustness of Trajectory Filtering}  \label{app:trstudy}
To enhance the robustness of trajectory selection, we propose a hybrid evaluation mechanism that fuses multi-trajectory cross-validation with vector similarity metrics. Specifically, this method first performs cross-validation analysis on multiple trajectories under the same task, while simultaneously calculating the cosine similarity of trajectory embeddings to quantify semantic consistency. We formulate a weighted hybrid evaluation function $E_{final}$, defined as follows:

\begin{equation}
    E_{final} = \alpha \cdot S_{cv} + (1 - \alpha) \cdot S_{sim}
    \label{eq:hybrid_eval}
\end{equation}

where $S_{cv}$ represents the cross-validation score, $S_{sim}$ denotes the vector similarity score, and the balancing weight $\alpha$ is set to 0.5.

\begin{table}[h]
\centering

\begin{tabular}{lcccc}
\toprule
\textbf{Dataset} & \textbf{MBPP }& \textbf{MathQA }&\textbf{MATH }&\textbf{FEVER}\\
\midrule
Accuracy   & 89.37\% &95.43\% & 92.05\% & 94.27\%\\

\bottomrule
\end{tabular}
\vspace{6pt}
\caption{The accuracy of trajectory selection.}
\label{tab:overhead}
\end{table}

To validate the effectiveness of this mechanism, we conducted tests on a constructed benchmark dataset. This dataset strictly controls data distribution, comprising 90\% valid trajectories and 10\% adversarial erroneous trajectories (simulating attack samples). Experimental results demonstrate that the proposed hybrid evaluation mechanism exhibits a strong capacity for suppressing malicious or erroneous trajectories while retaining high-confidence ones, achieving a filtering accuracy of \textbf{92.78\%}, which significantly outperforms single-metric evaluation schemes.

\section{Supplementary Ablation Study}  \label{app:ablation_study}
\begin{figure*}[ht]
\centering
\begin{minipage}[b]{0.326\textwidth}
\centering
\includegraphics[width=\linewidth]{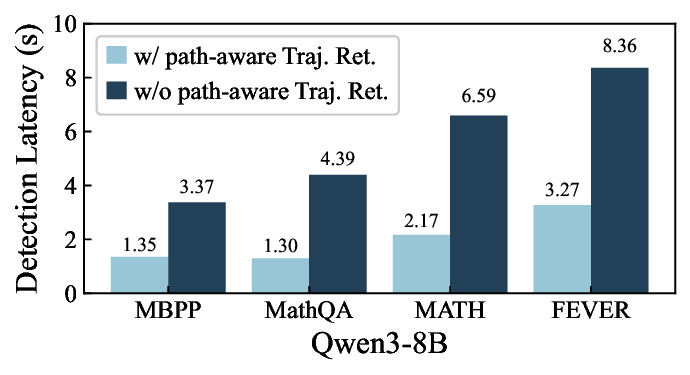}
\end{minipage}
\hfill
\begin{minipage}[b]{0.326\textwidth}
\centering
\includegraphics[width=\linewidth]{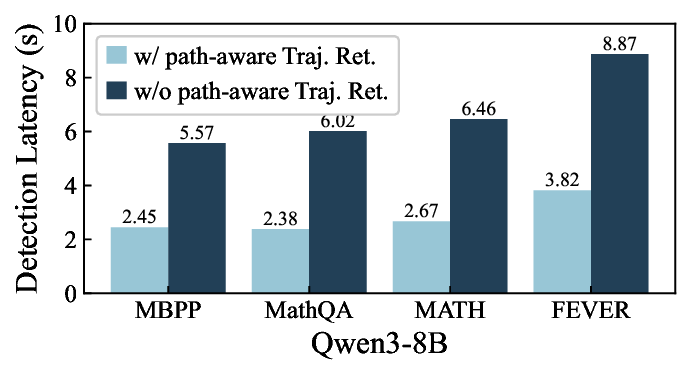}
\end{minipage}
\hfill
\begin{minipage}[b]{0.326\textwidth}
\centering
\includegraphics[width=\linewidth]{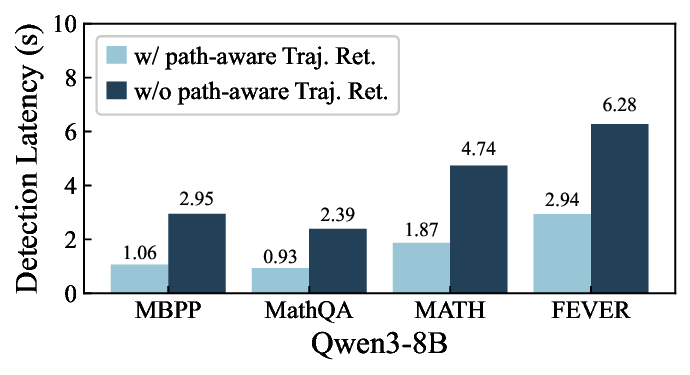}
\end{minipage}

\caption{Detection overhead when the path-aware trajectory retrieval method is enabled and disabled.}
\label{fig:abl_path_aware}
\end{figure*}
We evaluate the effectiveness of the proposed path-aware trajectory retrieval method (\S\ref{sec:retrieval}), which aims to reduce the number of trajectories required for workflow invalidation detection and thereby lower detection overhead. We compare the detection latency, including both communication overhead for retrieving trajectories from the remote repository and the verification latency on the GLM, with and without path-aware retrieval under workflows with varying structural complexity, where the number of branching nodes ranges from 2 to 6. We further conduct this evaluation across different GLMs used for workflow invalidation detection. The results are summarized in Figure~\ref{fig:abl_path_aware}.

Across all datasets and GLMs, the path-aware trajectory retrieval method consistently yields substantial latency reductions. On average, it reduces detection latency by 71\% compared to the baseline that retrieves trajectories without considering execution paths. Without path-aware retrieval, detection latency grows rapidly as workflow complexity increases, since trajectories associated with all possible paths are indiscriminately retrieved and verified. In the most complex setting, detection latency reaches up to 886.94 seconds, highlighting the poor scalability of path-agnostic retrieval.

In contrast, incorporating path-aware retrieval leads to significantly more stable scaling behavior. By restricting trajectory retrieval to the currently determined execution path, the detection latency remains bounded even as the workflow complexity increases. Across all evaluated settings, the latency with path-aware retrieval is consistently limited to within 382.25 seconds, demonstrating that the proposed method effectively decouples detection overhead from the total number of workflow paths.

We further observe that the benefits of path-aware retrieval are consistent across different GLMs, including Qwen3-8B, LLaMA3-8B, and Qwen2.5-7B. Although absolute latency varies across models due to differences in model size and verification cost, the relative latency reduction achieved by path-aware retrieval remains similar. This suggests that the efficiency gains stem from the algorithmic design of selective trajectory retrieval, rather than from model-specific optimizations.

\paragraph{Latency Overhead Evaluation.}
For the complexity of the path-aware strategy, consider an execution path $P$ with $|P|$ steps. For each step $v \in P$, the verifier retrieves $K$ aligned signed trajectories, verifies their signatures, and performs semantic consistency checking. Let $C_{\text{sig}}$ and $C_{\text{sem}}$ denote the costs of signature verification and semantic checking, respectively. The per-step cost is $O\!\left(K (C_{\text{sig}} + C_{\text{sem}})\right)$, yielding a total complexity of $O\!\left(|P| \cdot K (C_{\text{sig}} + C_{\text{sem}})\right)$.

As shown in table~\ref{tab:overhead} , to empirically validate the overhead, we deploy Qwen2.5-7B on two NVIDIA L20 GPUs (48GB each) and measure end-to-end latency on four datasets. Compared to inference-only execution, our method incurs an average latency increase of less than 14\%, demonstrating that the added verification overhead remains practically acceptable.

\begin{table}[h]
\centering

\begin{tabular}{lccc}
\toprule
\textbf{Dataset} & \textbf{Inference Only }& \textbf{Verification + Inference }& \textbf{Increase }\\
\midrule
MBPP   & 1.89s& 2.05s& 8.46\%\\
MathQA & 3.08s& 3.45s& 12.09\%\\
MATH   & 3.94s& 4.71s& 19.73\%\\
FEVER  & 6.95s& 8.04s& 15.62\%\\
\bottomrule
\end{tabular}
\vspace{6pt}
\caption{End-to-end latency comparison with and without verification.}
\label{tab:overhead}
\end{table}

\section{Limitations and Future Work}  \label{app:lf}
While the proposed workflow invalidation detection framework demonstrates strong effectiveness across multiple reasoning benchmarks, several limitations remain.

First, our framework relies on the assumption that the majority of trajectory-submitting agents are honest and that cryptographic signatures cannot be forged. Although this assumption is consistent with common settings in collaborative and federated systems, stronger adversaries that compromise a larger fraction of agents may still inject coordinated poisoned trajectories capable of influencing workflow abstraction and verification. Future work could explore Byzantine-resilient aggregation mechanisms and trust-aware trajectory weighting strategies to further improve robustness under large-scale adversarial participation.

Second, the current verification framework primarily focuses on workflow integrity and semantic consistency, but does not explicitly model long-term adaptive adversaries. An attacker may gradually evolve poisoning strategies to better mimic authentic execution trajectories while preserving semantic plausibility. Developing adaptive or continual verification mechanisms that evolve together with workflow repositories remains an important open problem.

Third, although the proposed path-aware trajectory retrieval strategy significantly reduces verification overhead, the runtime cost can still become substantial for highly branching workflows or large-scale agent ecosystems. In particular, verification latency depends on both the number of retrieved trajectories and the complexity of semantic consistency checking performed by the guided language model. More efficient lightweight verification models, hierarchical retrieval mechanisms, or compressed provenance representations may further improve scalability.

Fourth, our evaluation mainly considers workflow invalidation attacks operating at the repository and trajectory levels. Other attack surfaces in LLM-based agent systems, such as tool manipulation, retrieval corruption, memory poisoning, or environment-level adversarial interactions, are not explicitly addressed in this work. Future research could investigate unified security frameworks capable of jointly defending multiple components of agent ecosystems.

Finally, while our experiments cover diverse reasoning benchmarks including mathematical reasoning, factual verification, and code generation, the evaluation remains limited to controlled benchmark settings. Real-world deployment scenarios may involve substantially more dynamic workflows, heterogeneous agent behaviors, and noisy execution environments. Extending the proposed framework to real-world multi-agent platforms and production-scale workflow repositories is an important direction for future work.

Despite these limitations, we believe this work represents an important step toward trustworthy and verifiable shared workflow systems for LLM-based agents, and we hope it motivates further research on the security, reliability, and provenance of collaborative agent ecosystems.


\end{document}